\documentclass[aps,11pt]{revtex4}
\usepackage{epsfig}
\usepackage{amsmath}
\usepackage{bm}
\usepackage{times}
\usepackage{graphicx}
\begin{document}
\title{\huge Gravity Effects on Neutrino Masses in Split Supersymmetry}
\author{Marco Aurelio D\'\i az}
\author{Benjamin Koch}
\author{Boris Panes}
\affiliation{
{\small Departamento de F\'\i sica, Universidad Cat\'olica de Chile\\
Avenida Vicu\~na Mackenna 4860, Santiago, Chile 
}}
%%%%%%%%%%%%%%%%%%%%%%%%%%%%%%%%%%%%%%%%%%%%%%%%%%%%%%%%%%%%%%%%%%%%

\begin{abstract}
The mass differences and mixing angles of neutrinos can neither be explained by 
R-Parity violating split supersymmetry nor by flavor blind quantum gravity alone. 
It is shown that combining both effects leads, within the allowed parameter range, 
to good agreement with the experimental results. The atmospheric mass is generated 
by supersymmetry through mixing between neutrinos and neutralinos, while the solar 
mass is generated by gravity through flavor blind dimension five operators. 
Maximal atmospheric mixing forces the tangent squared of the solar angle to be 
equal to $1/2$. The scale of the quantum gravity operator is predicted within a 
$5\%$ error, implying that the reduced Planck scale should lie around the GUT
scale.
In this way, the model is very predictive and can be tested at future
experiments.
\end{abstract}

%%%%%%%%%%%%%%%%%%%%%%%%%%%%%%%%%%%%%%%%%%%%%%%%%%%%%%%%%%%%%%%%%%%%
\date{\today}
\maketitle

%%%%%%%%%%%%%%%%%%%%%%%%%%%%%%%%%%%%%%%%%%%%%%%%%%%%%%%%%%%%%%%%%%%%
\section{Introduction}
%%%%%%%%%%%%%%%%%%%%%%%%%%%%%%%%%%%%%%%%%%%%%%%%%%%%%%%%%%%%%%%%%%%

The existence of neutrino masses and mixing angles
is experimentally well confirmed and therefore the
theoretical understanding and description of those
quantities is one of the most urgent issues for particle
physics~\cite{NeutrinoReview}.
There are two frequently studied theoretical extensions
of the standard model of particle physics which also
were expected to explain the origin and the shape of the small
but non zero neutrino mass matrix. 
%Those two extensions are Supersymmetry and the possible
%existence non-renormalizable gravitational interactions that
%appear at a lowered Planck scale:

On the one hand there is Supersymmetry.
Although there is no experimental evidence for supersymmetry, it is
often invoked that the unification of gauge coupling constants 
at the GUT scale is an
indirect indication for supersymmetry. 
It has been pointed out that 
this gauge unification is achieved even if all scalar superpartners of
quarks and leptons are very heavy, introducing the scenario called
Split Supersymmetry (SS)~\cite{SPLIT,Bajc:2004hr}.
The original SS scenario includes R-Parity conservation, which guarantees 
the stability of the LSP and thus a Dark Matter candidate, and only one 
light Higgs doublet, which behaves like the SM Higgs. Despite the fact 
that models with R-Parity Violation (RpV) loose the LSP as a dark matter
candidate, they are studied because they provide a compelling mechanism for 
the generation of neutrino masses~\cite{Rparity-neutrinos}. 
Nevertheless, SS models with RpV
are incapable to produce the necessary neutrino masses, even
in next to leading order calculations 
\cite{Chun:2004mu,Diaz:2006ee,Davidson:2000uc}.

On the other hand there is the possible
existence of non-renormalizable gravitational interactions.
Those interactions could have
an influence on the neutrino sector of the standard
model~\cite{Barbieri:1979hc,Akhmedov:1992hh,Akhmedov:1992et,deGouvea:2000jp,
Dighe:2006sr,Vissani:2003aj}.
Although the standard Planck scale $M_P=1.2\times 10^{19}$~GeV
generates a solar mass which is too small to fit the experimental evidence,
a lowered Planck scale $M_f$ might in principle do the job.
However, gravitational interactions are expected to be ``flavor blind''
and it has been shown that therefore a purely gravitational neutrino 
sector is also incapable to explain
the existence of three different neutrino masses~\cite{Berezinsky:2004zb}.

In this paper we study the combined effect of 
the supersymmetrical and gravitational neutrino sector.
We find that the combination of both effects can
explain all present neutrino data. 
But this kind of model even leads to predictions
for the solar neutrino mixing angle $\sin^2 \theta_{sol}$ and
for the scale of the gravitational contribution.

%%%%%%%%%%%%%%%%%%%%%%%%%%%%%%%%%%%%%%%%%%%%%%%%%%%%%%%%%%%%%%%%%
\section{The neutrino mass matrix in R-parity violating SPLIT SUSY}
%%%%%%%%%%%%%%%%%%%%%%%%%%%%%%%%%%%%%%%%%%%%%%%%%%%%%%%%%%%%%%%%%%

In Split SUSY all scalars are very heavy, except for one Higgs
doublet~\cite{SPLIT}.
Since we are interested in the possibility to describe the neutrino masses 
in Split-SUSY, we need to consider R-Parity violating
interactions~\cite{Rparity}.
Knowing all relevant trilinear R-parity violating interactions one can 
calculate
the neutralino/neutrino mass matrix.
Integrating out the neutralinos from this matrix, one finds that the neutrino
mass matrix in flavor space is given by \cite{Diaz:2006ee}:
\begin{equation}
{\bf M}_\nu^{eff}=
%-m\,{\mathrm{M}}_{\chi^0}^{-1}\,m^T=
\frac{v^2}{4\det{M_{\chi^0}}}
\left(M_1 \tilde g_d^2 + M_2 \tilde g'^2_d \right)
\left[\begin{array}{cccc}
\lambda_1^2        & \lambda_1\lambda_2 & \lambda_1\lambda_3 \cr
\lambda_2\lambda_1 & \lambda_2^2        & \lambda_2\lambda_3 \cr
\lambda_3\lambda_1 & \lambda_3\lambda_2 & \lambda_3^2
\end{array}\right]\quad,
\label{treenumass}
\end{equation}
where the determinant of the neutralino mass matrix is:
\begin{equation}
\det{M_{\chi^0}}=-\mu^2 M_1 M_2 + \frac{1}{2} v^2\mu \left( 
M_1 \tilde g_u \tilde g_d + M_2 \tilde g'_u \tilde g'_d \right)
+\textstyle{\frac{1}{16}} v^4 
\left(\tilde g'_u \tilde g_d - \tilde g_u \tilde g'_d \right)^2\quad.
\label{detNeut}
\end{equation}
Here, $v$ is the vacuum expectation value of the light Higgs field,
$M_1$, $M_2$ are the gaugino masses. Further,
$\tilde {g}_{u,d}$ and $\tilde g'_{u,d}$ are the
trilinear couplings between the Higgs boson, the gauginos, and the higgsinos.
The parameters $\lambda_i\equiv a_i\mu+\epsilon_i$ \cite{Diaz:2006ee}
are related to the traditional BRpV parameters $\Lambda_i$
\cite{Nowakowski:1995dx} by $\Lambda_i=\lambda_i v_d$.
The $\epsilon_i$ are the parameters that mix higgsinos with
leptons, and $a_i$ are dimensionless parameters that mix gauginos with
leptons. 
Finally, $\mu$ is the higgsino mass.

This effective neutrino mass matrix ${\bf M}_\nu^{eff}$ has 
only one eigenvalue different from zero. As in the case of R-parity 
violation in the MSSM with bilinear terms, only one neutrino is massive. 
As it is explained in the literature~\cite{Chun:2004mu,Diaz:2006ee},
in SS it is not possible to explain the neutrino masses and mixing
using bilinear terms only.
Further allowing for trilinear couplings
makes it in principle possible to obtain solar neutrino masses and mixing.
However this possibility was not discussed here, because
it only works for a very special choice of parameters with
an undesired hierarchical structure among the trilinear
couplings~\cite{Chun:2004mu}. Without this hierarchical structure, the 
trilinear couplings become irrelevant because they contribute through loops 
of sparticles that have a mass of the order of the split supersymmetric
scale $\widetilde m$, and therefore, decoupled.

%%%%%%%%%%%%%%%%%%%%%%%%%%%%%%%%%%%%%%%%%%%%%%%%%%%%%%%%%%%%%%%%%%
\section{The neutrino mass matrix from low scale gravity effects}
%%%%%%%%%%%%%%%%%%%%%%%%%%%%%%%%%%%%%%%%%%%%%%%%%%%%%%%%%%%%%%%%%%
 
It is widely assumed that the unknown 
quantum gravity Lagrangian can be expanded to low 
energies.
In flavor space 
this expansion can give rise to a non-renormalizable term in the
Lagrangian of the type
\begin{equation}\label{QGoperator}
{\cal L} \sim \frac{\tilde{\lambda}_{ij}}{M_P}\overline\psi_i \psi_j 
\phi^2\quad.
\end{equation}
Here, $\psi_i$ and $\phi$ are the lepton and Higgs fields respectively.
In an idealized model 
the flavor mixing coefficients $\tilde{\lambda_{ij}}$ can all be
taken to be of order
one~\cite{Barbieri:1979hc,Akhmedov:1992hh,Akhmedov:1992et,deGouvea:2000jp,
Berezinsky:2004zb,Dighe:2006sr}.
When the Higgs acquires a vacuum expectation value, 
a neutrino mass is generated
\begin{equation}
m_\nu \sim {\cal O}(1)\frac{v^2}{M_P}\sim {\cal O}(10^{-6})
\,\,{\mathrm{eV}}\quad,
\end{equation}
with $v$ the electroweak vacuum expectation value. 
This type of contribution to the neutrino mass matrix
has also been explored in \cite{Joshipura:1998qn}.
Since this neutrino mass
term is too small, we look into the possibility that the true Planck 
scale $M_f$ is actually much lower than $M_P$.
Such a lowered Planck scale $M_f$ is an intrinsic prediction
of models with compact extra dimensions.
In some of those models the experimentally allowed Planck scale $M_f$ can be
as low as one TeV \cite{Antoniadis:1998ig,Randall:1999vf,Randall:1999ee}.
However, a TeV gravity scale in operators like in eq.~(\ref{QGoperator})
meets strong constraints from precision measurements such as 
$\mu \rightarrow e \gamma$ \cite{Berezhiani:1998wt}. This can be met 
by imposing
additional symmetries or by admitting that $M_f$ might not
be so extremely small
\begin{equation}
\label{inequality}
1\,{\mathrm{TeV}}\ll M_f<M_P\quad.
\end{equation}
Several
papers~\cite{Barbieri:1979hc,Akhmedov:1992hh,Akhmedov:1992et,deGouvea:2000jp,
Berezinsky:2004zb,Dighe:2006sr} 
discuss contributions of an exact ``blindness'' model, 
where the part of the
neutrino mass matrix coming
from gravitational effects can be parameterized 
in flavor space as
\begin{equation}\label{muMatrix}
\Delta M_g^\nu = \mu_g
\left[\begin{array}{cccc}
1 & 1 & 1 \cr
1 & 1 & 1 \cr
1 & 1 & 1 
\end{array}\right]\quad,
\end{equation}
where
\begin{equation}\label{eq_mugestimate}
\mu_g \sim {\cal O}(1)\frac{v^2}{M_f}\sim {\cal O}(10^{-2})
\,\,{\mathrm{eV}}\quad.
\end{equation}
It should be to noticed
that such an exact ``blindness'' model
does not imply base independence in flavor space.
This means that only when written in the flavor base 
(defined by a diagonal mass matrix of the charged Leptons),
the matrix (\ref{muMatrix}) takes its symmetric form.
Further, such a model
can only give direct predictions 
for the UPMNS matrix (\ref{UPMNS2})
if the relation of flavor basis to the mass basis 
is defined for the charged leptons as well.
Like in the standard model this relation is implicitly assumed
for the above contribution by taking both
basis to coincide~\cite{Dighe:2006sr}.

In order to explain the solar mass scale the fundamental 
scale has to be 
$M_f\sim {\cal O}(10^{15})\,\,{\mathrm{GeV}}$, which
is in good agreement with the inequality in eq.~(\ref{inequality}).
However, the matrix in eq.~(\ref{muMatrix}) has only one eigenvalue 
different from zero. Therefore such a scenario can
not account for both solar and atmospheric neutrino mass
splittings.

%%%%%%%%%%%%%%%%%%%%%%%%%%%%%%%%%%%%%%%%%%%%%%%%%%%%%%%%%%%%%%%%%%
\section{The neutrino mass matrix from R-parity violating SPLIT SUSY and 
low scale quantum gravity}
%%%%%%%%%%%%%%%%%%%%%%%%%%%%%%%%%%%%%%%%%%%%%%%%%%%%%%%%%%%%%%%%%%

Now we will discuss the possibility that both, R-parity violating 
Split Susy and low scale quantum gravity operators like 
eq.~(\ref{QGoperator}) contribute to the Lagrangian. 
In this case, two terms can contribute to the neutrino mass matrix.
Since the mass terms
in eqs.~(\ref{treenumass}) and (\ref{muMatrix})
are both formulated in flavor space one can write the total effective neutrino
mass matrix as,
\begin{equation}
\label{eq_mixmod}
M_g^{\nu ij}=A \lambda^i\lambda^j+\mu_g \quad,
\end{equation}
where we defined the factor in front of the mass matrix
in equation (\ref{treenumass}) as $A$.
As defined above, the neutrino mass matrix does not
contain CP violating phases.
In this case, if $A>0$ and $\mu_g>0$ the three neutrino masses are
\begin{eqnarray}
m_{\nu_1} &=& 0
\nonumber\\
m_{\nu_2} &=& \frac{1}{2}\left( A|\vec\lambda|^2+
3\mu_g\right)-\frac{1}{2}\sqrt{
\left( A|\vec\lambda|^2+3\mu_g\right)^2-
4A\mu_g|\vec v\times\vec\lambda|^2}
\\
m_{\nu_3} &=& \frac{1}{2}\left( A|\vec\lambda|^2+
3\mu_g\right)+\frac{1}{2}\sqrt{
\left( A|\vec\lambda|^2+3\mu_g\right)^2-
4A\mu_g|\vec v\times\vec\lambda|^2}
\nonumber
\end{eqnarray}
where we have defined the auxiliary  vector $\vec v=(1,1,1)$.
In the approximation where the $\mu_g$ term is subdominant, the 
squared mass differences are given by,
\begin{eqnarray}\label{eq_msolapprox}
\Delta m^2_{sol}&=&(m_2^2-m_1^2)=\mu_g^2\frac{(\vec{v}\times
\vec{\lambda})^4}{\vec{\lambda}^4}+{\mathcal{O}}(\mu_g^3)\\
\label{eq_matmapprox}
%\delta m^2_{reac}&=&(m_3^2-m_1^2)=A \vec{\lambda}^4+2 A \mu_g
%(\vec{v}\vec{\lambda})^2+{\mathcal{O}}(\mu_g^3)\\
\Delta m^2_{atm}&=&(m_3^2-m_2^2)=A \vec{\lambda}^4+2 A \mu_g
(\vec{v}\vec{\lambda})^2+{\mathcal{O}}(\mu_g^3)
\end{eqnarray}
This shows that for a very small $\mu_g$, 
%the atmospheric and the reactor
%mass square difference have the same behavior. 
the atmospheric mass scale is controlled by the parameter $A$ and the solar 
mass scale is controlled by $\mu_g$.
In this approximation, the eigenvectors are equal to,
\begin{eqnarray}
\vec v_1 &=& \frac{\vec v\times\vec\lambda}{|\vec v\times\vec\lambda|}
\nonumber\\
\vec v_2 &=& \frac{\vec\lambda\times(\vec v\times\vec\lambda)}
{|\vec\lambda\times(\vec v\times\vec\lambda)|}+{\cal O}(\mu_g)
\label{eigenvectors}\\
\vec v_3 &=& \frac{\vec\lambda}{|\vec\lambda|}+{\cal O}(\mu_g)
\nonumber
\end{eqnarray}
and the matrix $U_{PMNS}$ is formed with the eigenvectors in its columns.
The neutrino mixing matrix is defined as
\begin{equation}
U_{PMNS}=U_{32}U_{31}U_{21}\quad,
\end{equation}
where $U_{32}$ is a rotation matrix around the axis one (with cyclic 
permutations for the other two).

At this point it is instructive to have a closer
look at the matrix $U_{PMNS}$. This
matrix 
appears in the Lagrangian of the leptonic charged current interactions
such that,
\begin{equation}
{\cal L}\owns -\frac{g}{\sqrt{2}}
\big(\bar e_L^+,\bar\mu_L^+,\bar\tau_L^+\big)
W_\mu^-\gamma^\mu U_{PMNS} 
\left( \begin{matrix} \nu_1 \cr \nu_2 \cr \nu_3 \end{matrix} \right)
\end{equation}
where fermions are in the mass eigenstate basis. In the most general 
situation, both the charged lepton and neutrino mass matrices are 
non-diagonal in the interaction basis,
\begin{equation}
{\cal L}\owns -\big(\bar e'^+_L,\bar\mu'^+_L,\bar\tau'^+_L \big) M_\ell
\left( \begin{matrix} e'^-_L \cr \mu'^-_L \cr \tau'^-_L \end{matrix} \right)
-\big(\bar\nu'^c_1,\bar\nu'^c_2,\bar\nu'^c_3 \big) M_\nu
\left( \begin{matrix} \nu'_1 \cr \nu'_2 \cr \nu'_3 \end{matrix} \right)
\end{equation}
where the prime on the fermion fields denote the interaction basis. The
two mass matrices are diagonalized by $V_\ell$ and $V_\nu$,
\begin{equation}
V_\ell^\dagger M_\ell V_\ell = {\mathrm{diag}}(m_e,m_\mu,m_\tau)
\,,\qquad
V_\nu^\dagger M_\nu V_\nu = {\mathrm{diag}}(m_{\nu_1},m_{\nu_2},m_{\nu_3})
\end{equation}
and the $U_{PMNS}$ matrix becomes, 
\begin{equation}\label{UPMNS2}
U_{PMNS}=V_\ell^\dagger V_\nu,
\end{equation} 
where 
we have assumed CP conservation. Due to the Majorana nature of neutrinos, 
$U_{PMNS}$ depends on three mixing angles and three Majorana phases. We 
assume the latter to be zero.
In general, a simultaneous diagonalization of $M_\nu$ and $M_\ell$
is necessary to find $U_{PMNS}$.
In principle there might be also 
off-diagonal contributions to the matrix
$M_\ell$ coming from the ``flavor blind'' gravitational terms in
eq.~(\ref{QGoperator})
(R-parity violating supersymmetric intercations do not
modify $V_\ell$).
However, those contributions (if they are present)
can be depreciated in our model since
they only can produce corrections to the diagonal
mass matrix of the charged leptons of the order
$\mu_g/m_e\sim10^{-8}$. This makes $V_\ell$ equal to unity up to
terms of the order of $\mu_g/m_e$.
In order to have a predictive model
we do not consider any other beyond the standard
model contributions, like non-diagonal terms
in the Yukawa couplings of the charged leptons.
Therefore, one can
identify $U_{PMNS}$ with $V_\nu$. 
It is in this basis that we include the
``flavor blind'' contribution from gravity given in eq.~(\ref{muMatrix}).

%%%%%%%%%%%%%%%%%%%%%%%%%%%%%%%%%%%%%%%%%%%%%%%%%%%%%%%%%%%%%%%%%%%%
\section{Results and Predictions}
%%%%%%%%%%%%%%%%%%%%%%%%%%%%%%%%%%%%%%%%%%%%%%%%%%%%%%%%%%%%%%%%%%%

In this section we present some numerical results, where we find
eigenvalues and eigenvectors of the $3\times3$ effective neutrino mass 
matrix using numerical methods. With them, we find neutrino mass 
differences and mixing angles, and compare them with values from 
experimental measurements. The solar and atmospheric mass differences, 
as well as the solar and atmospheric mixing angles, have been measured 
in several experiments.
\begin{table}
\begin{center}
\caption{Experimental measurements for neutrino parameters}
\bigskip
\begin{minipage}[t]{0.8\textwidth}
\begin{tabular}{|c|c|c|c|}
\hline
Observable $o_i$ & Mean value $\bar{o}_i$ &  $3\sigma_i$ variance & 
Units\\ 
\hline
$\Delta m_{atm}^2$ & $2.35 \times 10^{-3}$ & $0.95 \times 10^{-3}$ & 
eV${}^2$ \\ 
$\Delta m_{sol}^2$ & $8.15 \times 10^{-5}$ & $0.95 \times 10^{-5}$ & 
eV${}^2$ \\ 
$\sin^2 \theta_{atm}$ & $0.51$ & $0.17$ & - \\ 
$\sin^2 \theta_{sol}$ & $0.305$ & $0.075$ & - \\ 
\hline
\end{tabular}
\end{minipage}
\end{center}
\end{table}
We use the results of the combined analysis in ref.~\cite{Maltoni:2004ei}, 
given in Table I.
\begin{table}
\begin{center}
\caption{Experimental upper bounds for neutrino parameters}
\bigskip
\begin{minipage}[t]{0.8\textwidth}
\begin{tabular}{|c|c|c|}
\hline
Observable $o_i$ & $3\sigma_i$ upper bound & Units \\ 
\hline
$\sin^2 \theta_{reac}$ & $0.047$ & - \\ 
$m_{\beta\beta}$ & $0.84$ & eV \\
\hline
\end{tabular}
\end{minipage}
\end{center}
\end{table}
In addition, upper bounds have been obtained for the reactor angle 
\cite{Maltoni:2004ei}, and for the neutrino less double beta decay 
mass parameter $m_{\beta\beta}$. 
The value for the neutrino less double beta decay parameter is quite
precisely determined in our setting. Our model allows 
values of $0.001$~eV$<m_{\beta\beta}<0.005$~eV which is 
two orders of magnitude below
the current experimental bound.

We scan the parameters space, varying randomly the parameters $\lambda_i$,
$A$, and $\mu_g$, and calculate the goodness of the model, represented by
eq.~(\ref{eq_mixmod}), with
\begin{eqnarray}
\chi^2=\sum_i^6 \frac{(o_i-\bar{o}_i)^2}{(3 \sigma_i)^2} \quad.
\end{eqnarray}
where we have assigned a null mean value to the two parameters in 
Table II for which only upper bounds are known. 
Positive and negative solutions for $A$ were found.
A typical solution for negative $A$ is
\begin{eqnarray}
\label{benchmark}
& \lambda_1=0.0148\, {\mathrm{GeV}^2}\,,\quad 
  \lambda_2=0.0822\, {\mathrm{GeV}^2}\,, &
  \lambda_3=-0.0712\,{\mathrm{GeV}^2}\,, \cr
& A=-4.10\,{\mathrm{eV}}/{\mathrm{GeV}}^4\,, & \mu_g=0.003\,{\mathrm{eV}}
\end{eqnarray}
with $\chi^2=0.02$.
This shows that it is possible to find solutions for the combined model 
which are in ($3\sigma$) agreement with every single neutrino 
observable.
\begin{table}
\begin{center}
\caption{Predictions for neutrino parameters}
\bigskip
\begin{minipage}[t]{0.8\textwidth}
\begin{tabular}{|c|c|c|c|}
\hline
Observable & Value & Units\\ 
\hline
$\Delta m_{atm}^2$ & $2.39 \times 10^{-3}$ & eV${}^2$ \\ 
$\Delta m_{sol}^2$ & $7.74 \times 10^{-5}$ & eV${}^2$ \\ 
$\sin^2 \theta_{atm}$  & $0.596$  & - \\ 
$\sin^2 \theta_{sol}$  & $0.321$ & - \\ 
$\sin^2 \theta_{reac}$ & $0.023$ & - \\ 
$m_{\beta\beta}$ & $0.0039$ & eV \\
\hline
\end{tabular}
\end{minipage}
\end{center}
\end{table}
The prediction for the neutrino parameters in this case are summarized in 
Table III.

Based on the same $3\sigma$-limits scan, we study how the parameters of
this model are constrained by the experimental data. A key ingredient 
of this model is that it offers a possible way to measure $\mu_g$ for 
``flavor blind'' gravity effects. Indeed, an interesting prediction is that 
the allowed values of the mass parameter $\mu_g$ are strongly 
constrained. 
\begin{figure}[ht]
\includegraphics[width=10cm]{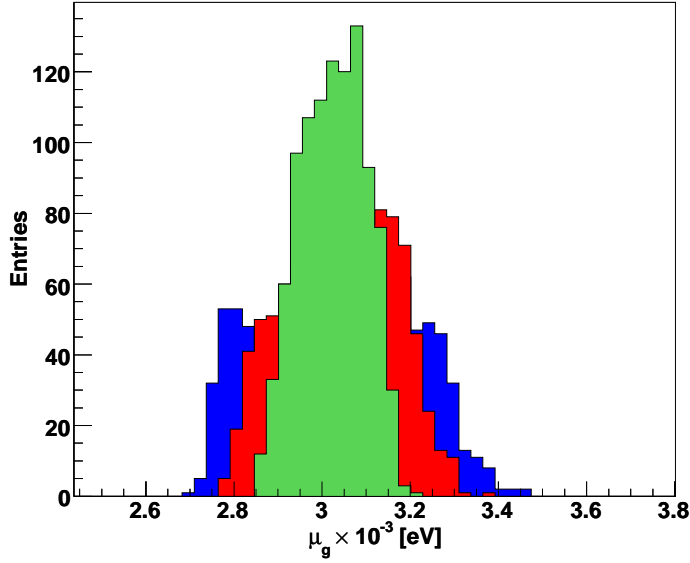}
\caption{Prediction for the mass parameter $\mu_g$ among models
that satisfy all the experimental constraints.
\label{events_ug}}
\end{figure}
In Fig.~\ref{events_ug} we show the frequency of occurrence of each value 
of $\mu_g$ among the selected models. This model predicts values centered 
around $\mu_g=3\times 10^{-3}$ eV with a 5\% error, as we can see from the 
green region defined by $\chi^2<1$. For comparison we show also the red 
region for $\chi^2<2$, and the blue region for $\chi^2<3$. According to 
eq.~(\ref{eq_mugestimate}), this implies a prediction for the true
Planck scale given by $M_f\approx 2\times 10^{16}$ GeV, which is 
remarkably similar to the GUT scale.

\begin{figure}[ht]
\begin{tabular}{cc}
\includegraphics[width=7cm]{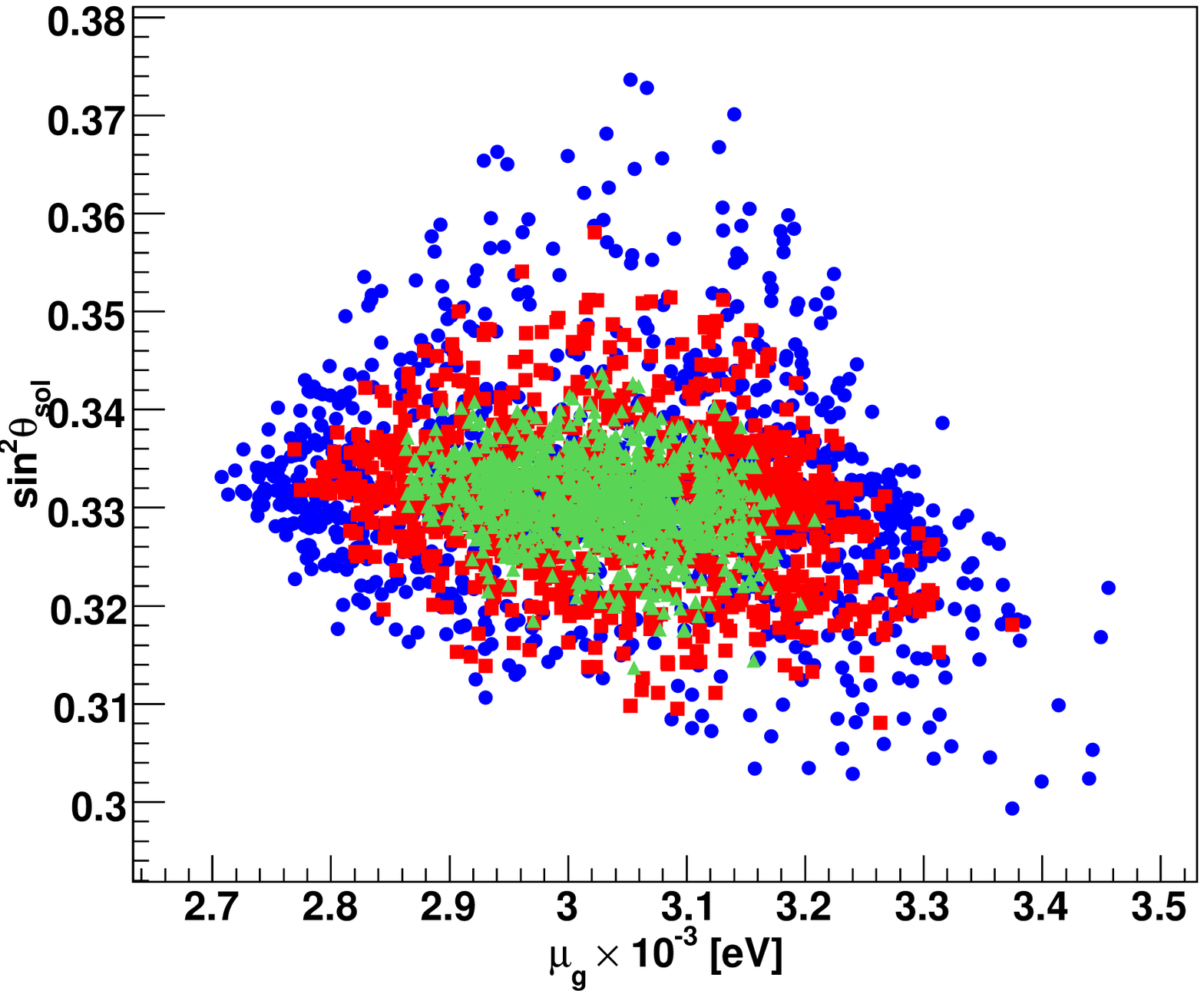}&
\includegraphics[width=7cm]{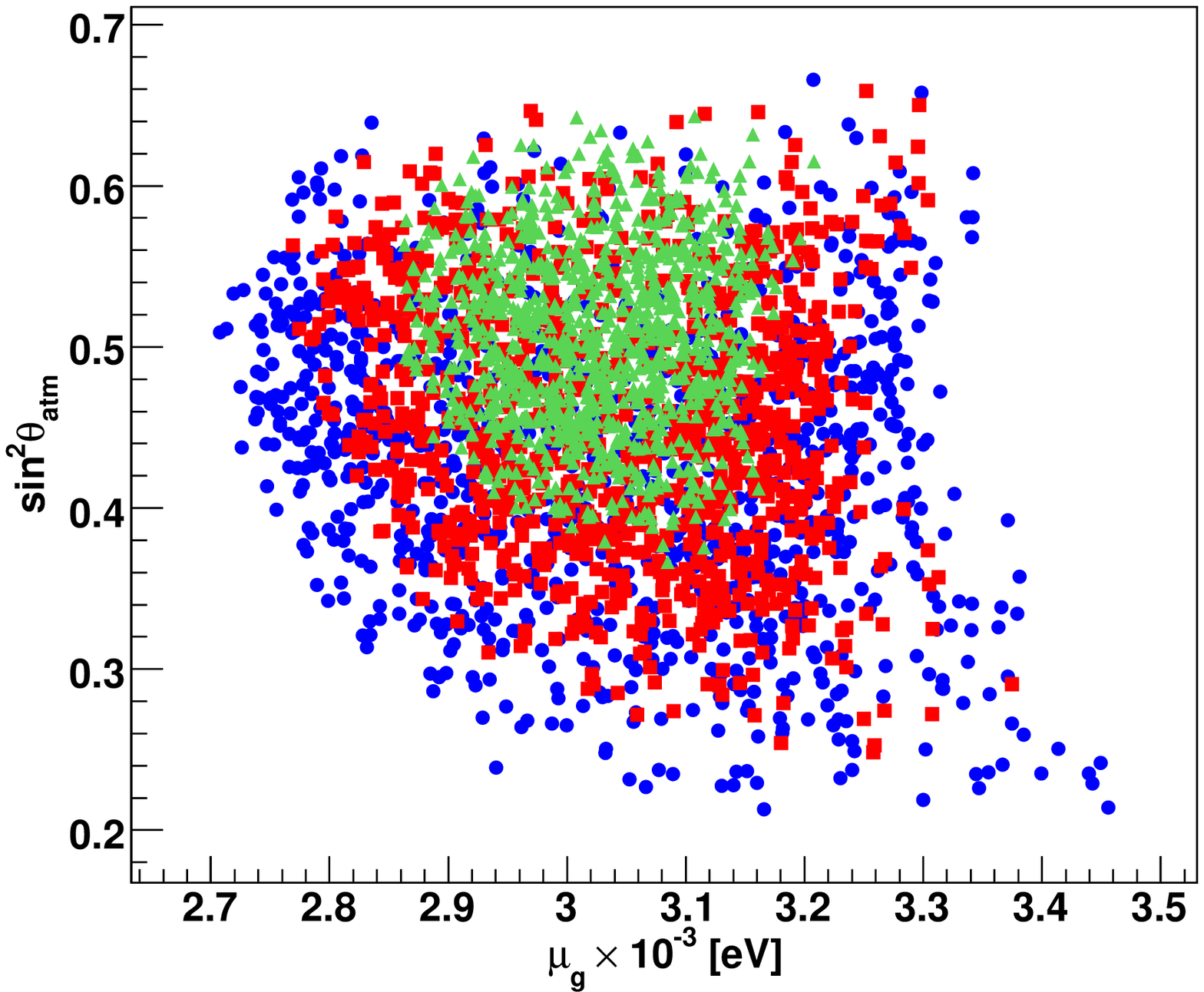}
\end{tabular}
\caption{Sections of the parameter space showing 
$\sin^2 \theta_{sol}$ and $\sin^2 \theta_{atm}$ as a function of $\mu_g$, with 
$\chi^2<1,2,3,$ for green triangles, red squares, and blue circles 
respectively. 
\label{angles_mug}}
\end{figure}
The same scan is shown in Fig.~\ref{angles_mug} where we see the 
neutrino mixing angles in correlation with the gravitational neutrino 
mass parameter $\mu_g$. Concentrating on the green triangles 
($\chi^2<1$), we see that the values of $\sin^2\theta_{sol}$ (left 
frame) are centered around 0.33, which is in the larger side of the 
experimentally allowed window. On the other hand, the values of
$\sin^2\theta_{atm}$ are centered around one half, which is right in the
center of the experimental window. For comparison we show also points
of parameter space less favored by experimental data, with red squares
corresponding to $\chi^2<2$, and blue circles to $\chi^2<3$.

Since the reactor angle satisfy,
\begin{equation}
\sin^2\theta_{reac}=(v_{3,1})^2=
\frac{\lambda_1^2}{|\vec\lambda|^2}<0.047
\end{equation}
where $v_{3,1}$ is the first component of the third eigenvector in
eq.~(\ref{eigenvectors}), and the quoted upper bound corresponds to 
the one given in Table II, we need 
$\lambda_1^2\ll\lambda_2^2+\lambda_3^2$. Neglecting the value of 
$\lambda_1$ in front of $\lambda_2$ and $\lambda_3$ we obtain,
\begin{eqnarray}
\tan^2\theta_{atm} &=& \left(\frac{v_{3,2}}{v_{3,3}}\right)^2 =
\frac{\lambda_3^2}{\lambda_2^2} = 1
\nonumber\\
\tan^2\theta_{sol} &=& \left(\frac{v_{2,1}}{v_{3,1}}\right)^2 =
\frac{\lambda_2^2+\lambda_3^2}{(\lambda_3-\lambda_2)^2} = \frac{1}{2}
\label{angles}
\end{eqnarray}
The numerical values for each parameter shown in eq.~(\ref{angles}) are 
obtained in the following scenario. Experimental results indicate 
$\sin^2\theta_{atm}=0.51\pm0.17$, where again we indicate the 
$3\sigma$ error. Maximal mixing is satisfied if we take
$\lambda_3^2=\lambda_2^2$. 
Choosing the sign $\lambda_3=-\lambda_2$ 
leads to the prediction for the solar angle indicated in the previous 
equation, $\sin^2\theta_{sol}=1/3$, which nicely agrees with the 
experimental result $\sin^2\theta_{sol}=0.305\pm0.075$.

\begin{figure}[ht]
\begin{tabular}{cc}
\includegraphics[width=7cm]{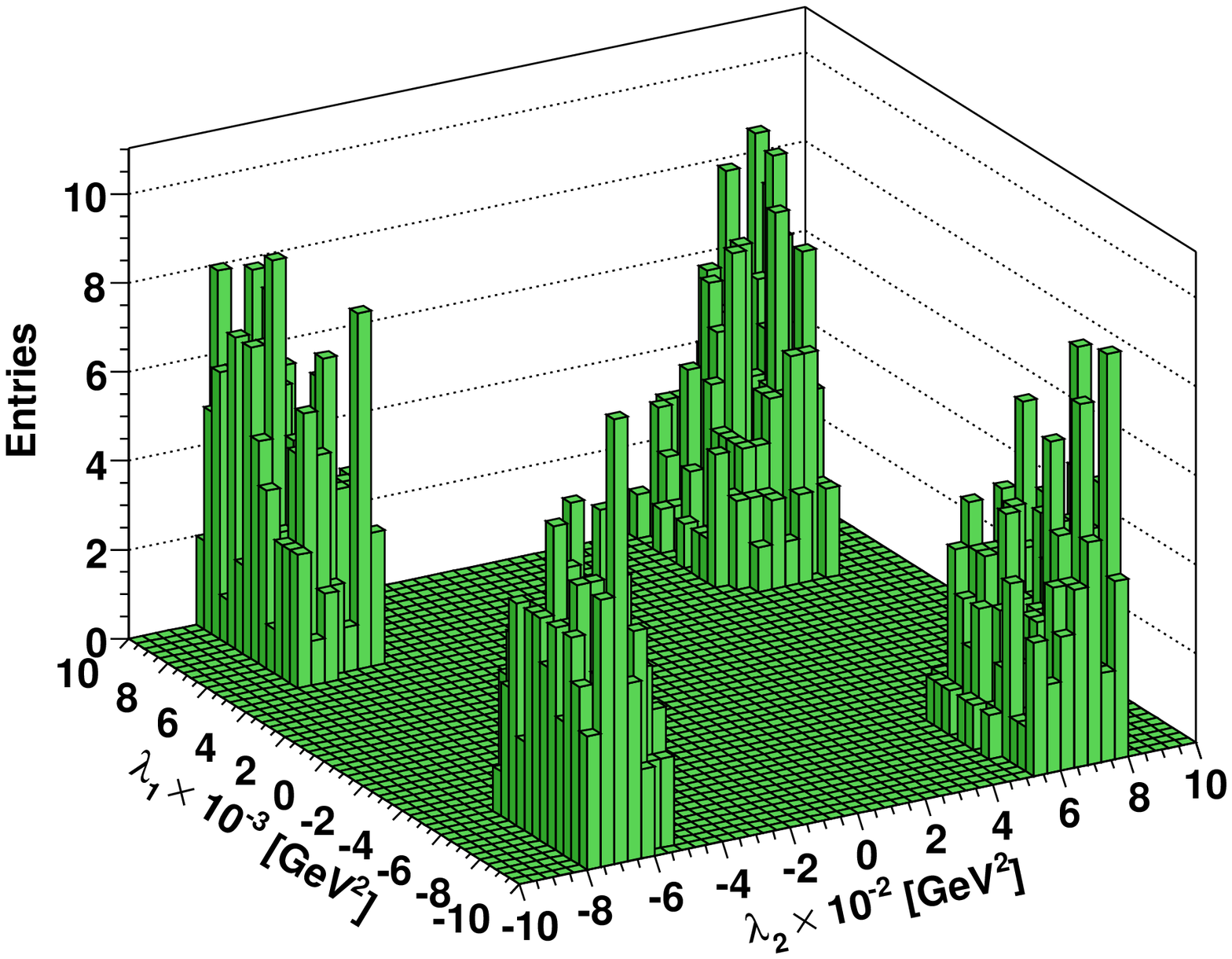}&
\includegraphics[width=7cm]{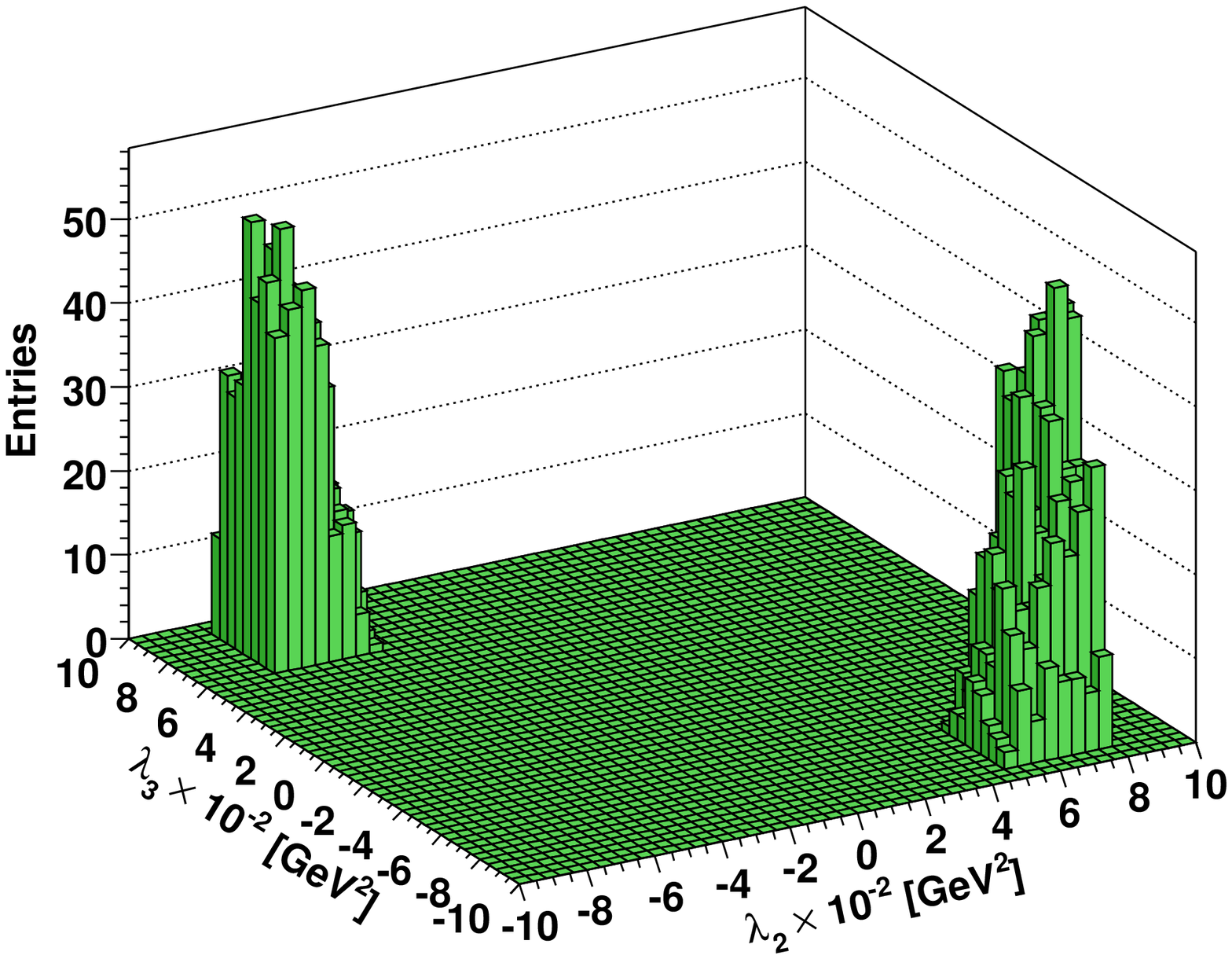}
\end{tabular}
\caption{Frequency of occurrence for the parameters $\lambda_i$ among
the models compatible with experimental results.
\label{lambdas}}
\end{figure}
As we will see now, this prediction is confirmed by the scan and the
numerical diagonalization of the neutrino mass matrix. In 
Fig.~\ref{lambdas} we plot the frequency of occurrence for each of the 
parameters $\lambda_i$ among the models in our scan which satisfy the
experimental constraints detailed in Tables I and II. We see in the left
frame that $\lambda_1$ is typically an order of magnitude smaller than 
the other two parameters $\lambda_2$ and $\lambda_3$, consistent with 
requirements from the reactor angle. 
Due to this hierarchy in the parameters, our model
has some interesting common features with the
tri-bi-maximal mixing model discussed in \cite{Dighe:2006sr}.
In the right frame we see that 
models consistent with experiments need $\lambda_3=-\lambda_2$, leading 
to the prediction $\sin^2\theta_{sol}=1/3$ as explained earlier.
\begin{figure}[ht]
\includegraphics[width=10cm]{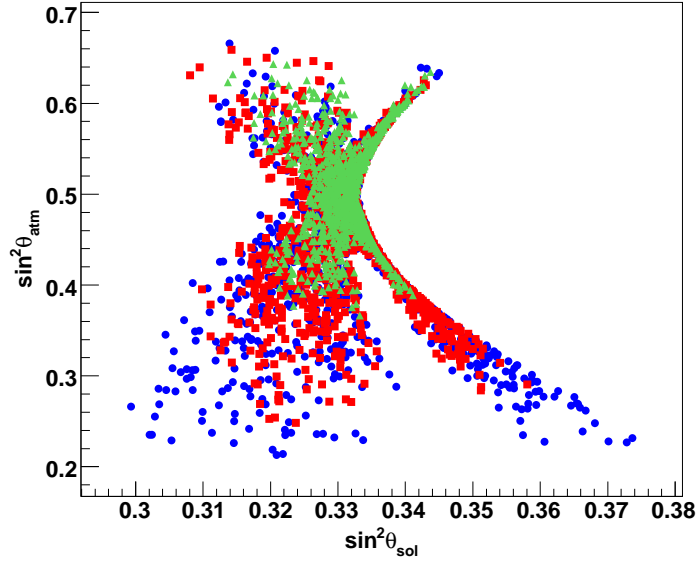}
\caption{Section of the parameter space showing 
$\sin^2 \theta_{atm}$ versus $\sin^2 \theta_{sol}$, with $\chi^2<1,2,3$ for 
green triangles, red squares, and blue circles respectively. 
\label{angle_atm_angle_sol}}
\end{figure}
The prediction $\tan^2\theta_{23}=1\Rightarrow \tan^2\theta_{sol}=1/2$,
is observed also in our scan. In Fig.~\ref{angle_atm_angle_sol} the 
relation between the neutrino mixing parameters $\sin^2 \theta_{atm}$
and $\sin^2 \theta_{sol}$ is shown. From this plot one sees that the 
model can only deliver a good agreement with all experimental bounds, 
if $0.313<\sin^2 \theta_{sol}<0.342$. This very strong constraint on 
$\sin^2 \theta_{sol}$ (remember that the experimentally allowed region 
is $0.23<\sin^2 \theta_{sol}<0.38$) gets even smaller if the atmospheric 
mixing parameter $\sin^2 \theta_{atm}$ is taken at its central value
of $0.51$, as seen in Fig.~\ref{angle_atm_angle_sol}. In this case one 
finds that the model predicts 
$0.325<\sin^2 \theta_{sol}<0.334$. With those small theoretical 
uncertainties an improved measurement of $\sin^2 \theta_{sol}$ would 
already allow to confirm the model prediction or otherwise rule out 
this model.

\begin{figure}[ht]
\begin{tabular}{cc}
\includegraphics[width=7cm]{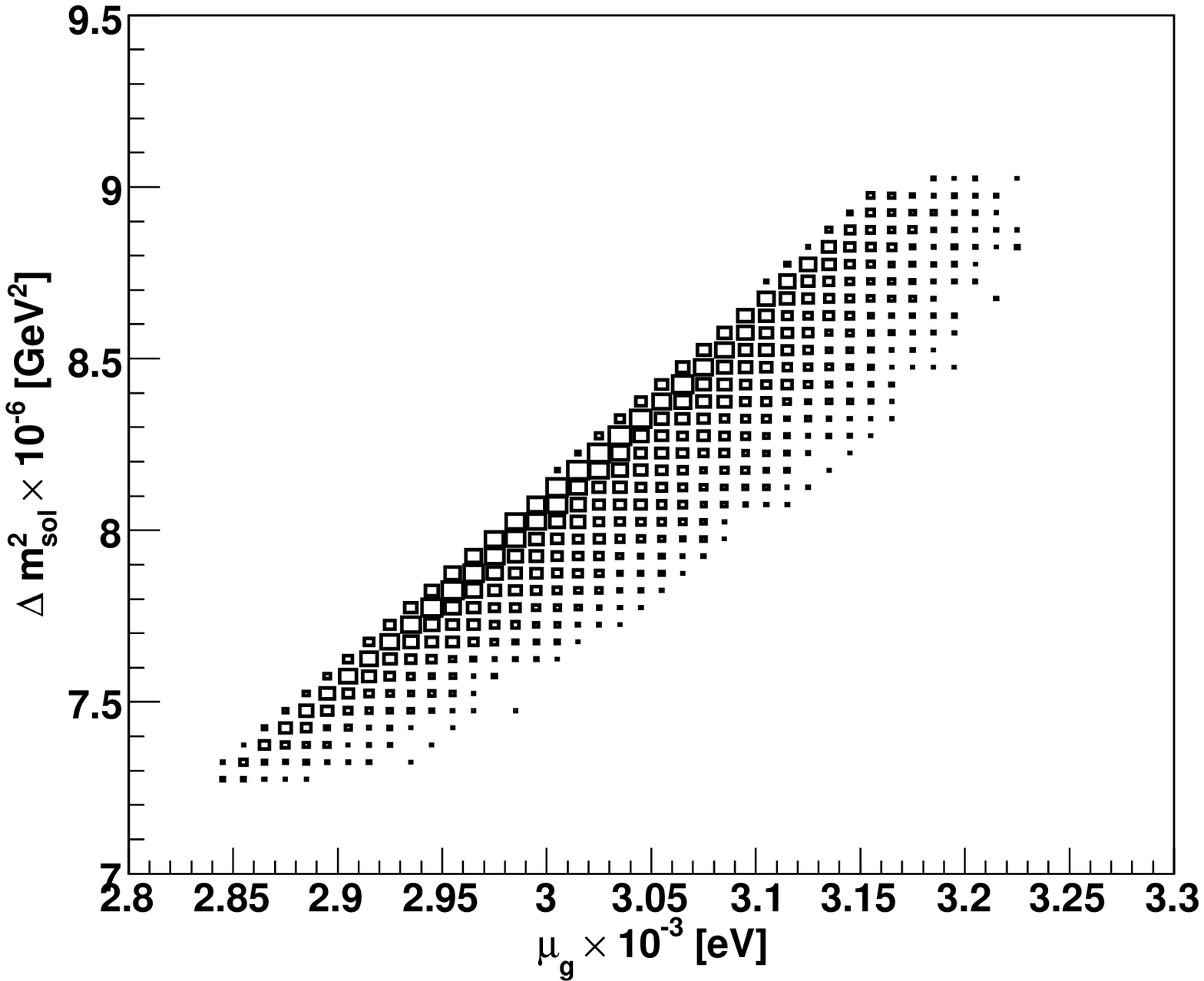}&
\includegraphics[width=7cm]{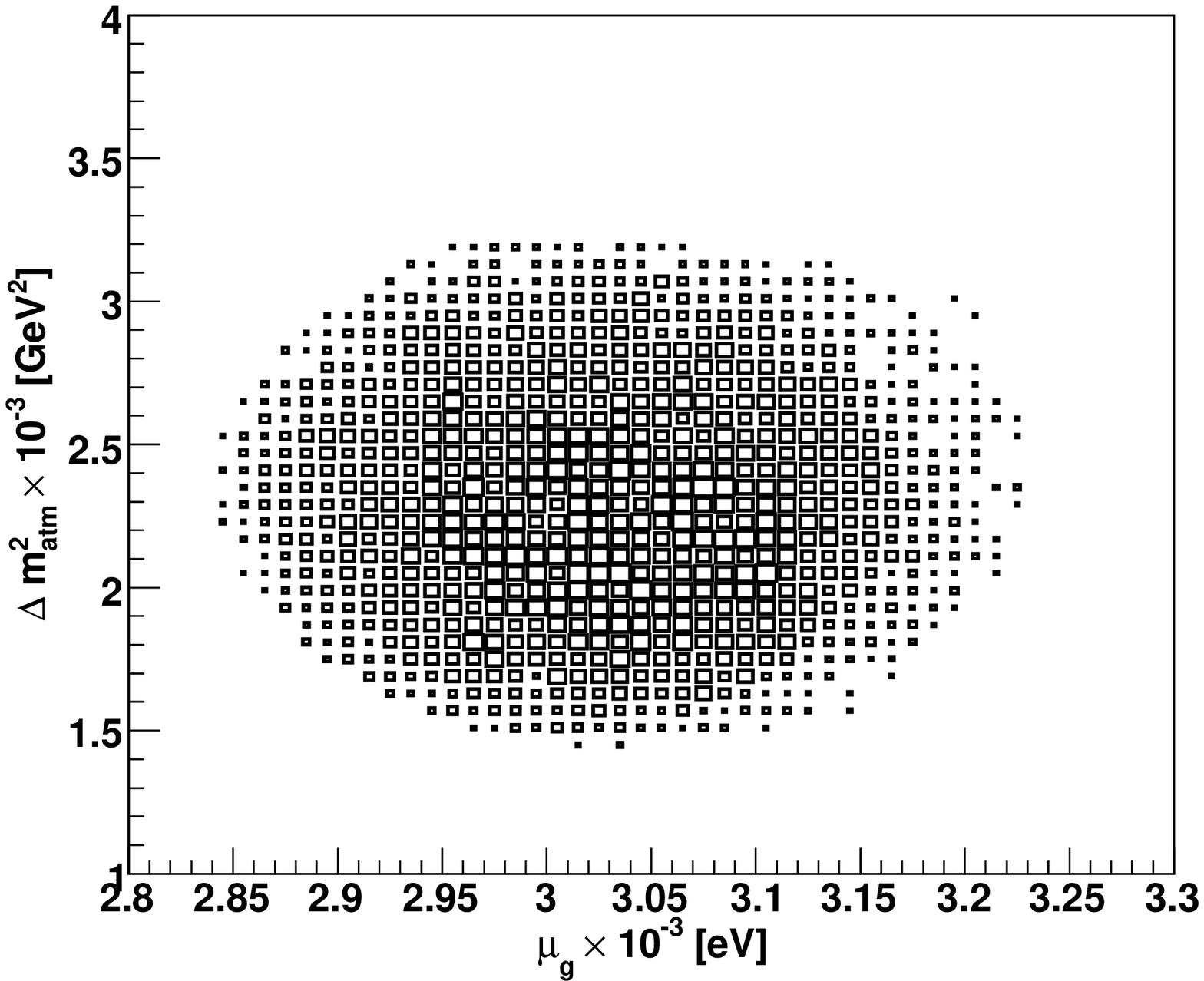}
\end{tabular}
\caption{Sections of the parameter space showing $\Delta m^2_{sol}$ and 
$\Delta m^2_{atm}$ as a function of $\mu_g$ with $\chi^2<1$.
\label{masses_mug}}
\end{figure}
The neutrino mass differences $\Delta m^2_{sol}$ and $\Delta m^2_{atm}$
are shown in Fig.~\ref{masses_mug}. The left frame shows that a growing 
$\mu_g$ gives a growing $\Delta m^2_{sol}$, which can be understood by 
comparing to the approximation in eq.~(\ref{eq_msolapprox}), that is, the 
$A$ term dominates over the $\mu_g$ term in eq.~(\ref{eq_mixmod}). In the
right frame we have the atmospheric mass difference $\Delta m^2_{atm}$
as a function of the gravitational neutrino mass parameter $\mu_g$, where
there is no obvious dependence because the atmospheric mass difference is 
dominated by the $A$ term. In both mass differences though, we see that 
the predictions lie nicely at the center of the experimental window.

It is also instructive to study how the scale of the supersymmetric term 
$A$ and the scale of the gravitational term $\mu_g$ are pinned 
down by the individual experimental constraints in Table I. Starting from 
the benchmark point in eq.~(\ref{benchmark}) and varying with respect to 
$A$ and $\mu_g$ we plot the allowed regions for every observable and the 
intersection of those regions.
\begin{figure}[ht]
\includegraphics[width=7cm]{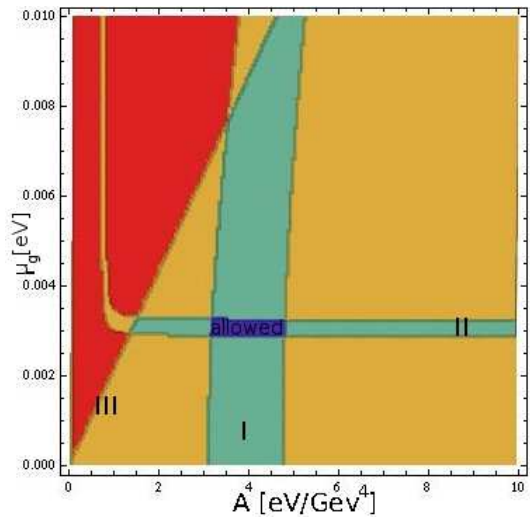}
\caption{Allowed regions in the positive $A$-$\mu_g$ plane.
\label{exclusionplot}}
\end{figure}
At the center of Fig.~\ref{exclusionplot} we have the zone in the $A$-$\mu_g$ 
plane which is consistent with experiments, and formed by the intersection of 
several regions. Region I is a vertical stripe where mainly the $A$ parameter 
is constrained by the $\Delta m_{atm}^2$ data. Region II is a horizontal 
stripe where $\mu_g$ is constrained by the $\Delta m_{sol}^2$ data. Note that 
this horizontal stripe turns into a vertical one after an eigenvalue crossing,
although this last branch is not allowed by the solar angle data. Finally, 
region III indicates the constraint from $\sin^2\theta_{sol}$, whose allowed 
region is at the right of the diagonal line, which breaks at the top also due 
to an eigenvalue crossing. Constraints from the other neutrino observables are 
not relevant for this figure. 

The main features from Fig.~\ref{exclusionplot}
can be understood by looking at the approximations in eqs.~(\ref{eq_msolapprox})
and (\ref{eq_matmapprox}), which show that for small values of $\mu_g$, the
atmospheric mass difference $\Delta m_{atm}^2$ is dominated by $A$, while the 
solar mass difference $\Delta m_{sol}^2$ is dominated by $\mu_g$. One sees that 
in the vicinity of this benchmark point, the other observables play a minor 
role in pinning down the scale parameters $A$ and $\mu_g$.

%%%%%%%%%%%%%%%%%%%%%%%%%%%%%%%%%%%%%%%%%%%%%%%%%%%%%%%%%%%%%%%%%%%%%%%%%%%%%%%

\section{Summary}

It is known that in Split Supersymmetry with R-Parity violation an atmospheric 
mass difference is generated at tree level, but one-loop contributions are
not enough to lift the symmetry of the effective neutrino mass matrix, thus
not being able to generate a solar mass difference. This problem is not solved 
by adding the, in principle always present, ``flavor blind'' couplings from
dimension 
five operators. The reason is that the Planck scale is too large to generate a 
solar mass difference large enough to be compatible with experiments. In this 
article we have shown that Split Supersymmetry with R-Parity violation, plus 
``flavor blind'' gravity effects with a reduced Planck scale, present in models
with 
compact extra dimensions, can be compatible with all data form neutrino 
experiments. The atmospheric mass difference is generated by supersymmetry with
a 
mixing between neutrinos and neutralinos, while the solar mass difference is 
generated by the ``flavor blind'' gravitational effects. This model predicts a
value 
for the gravitational term $\mu_g=3\times 10^{-3} \pm 5\%$ eV,
which 
corresponds to a reduced Planck scale $M_f\approx 2\times 10^{16}$ GeV. The fact 
that this reduced Planck scale is equal to GUT scale is a tantalizing result that 
may be related to gauge coupling unification of all four forces. In addition, the 
solar mixing angle is predicted to satisfy $0.313<\sin^2 \theta_{sol}<0.342$. We 
show also that a maximal atmospheric mixing $\sin^2\theta_{atm}=1/2$ implies 
$\sin^2\theta_{sol}=1/3$, which agrees with the implications from our parameter 
scan $0.325<\sin^2 \theta_{sol}<0.334$, when we adopt the central value 
$\sin^2\theta_{atm}=0.51$. In this way, the model not only reproduce the 
experimental results but it is also predictive and, therefore, can be falsified
by future experiments.

%%%%%%%%%%%%%%%%%%%%%%%%%%%%%%%%%%%%%%%%%%%%%%%%%%%%%%%%%%%%%%%%%%
\begin{acknowledgments}
{\small 
We are indebted to Dr.~Pavel Fileviez-P\'erez for his insight in the early
stages of this work.
The work of M.A.D. was partly funded by Conicyt-PBCT grant 
No. ACT028 (Anillo Centro de Estudios Subat\'omicos), and by 
Conicyt-PBCT grant ACI35. B.K. was funded by Conicyt-PBCT grant PSD73.
B.P. was funded by Conicyt's ``Programa de Becas de Doctorado''.}  
\end{acknowledgments}
%%%%%%%%%%%%%%%%%%%%%%%%%%%%%%%%%%%%%%%%%%%%%%%%%%%%%%%%%%%%%%%%%

%\newpage

%%%%%%%%%%%%%%%%%%%%%%%%%%%%%%%%%%%%%%%%%%%%%%%%%%%%%%%%%%%%%%%%%%
%\appendix
%%%%%%%%%%%%%%%%%%%%%%%%%%%%%%%%%%%%%%%%%%%%%%%%%%%%%%%%%%%%%%%%%%%%%%%%%

%%%%%%%%%%%%%%%%%%%%%%%%%%%%%%%%%%%%%%%%%%%%%%%%%%%%%%%%%%%%%%%%%%%%%%%%%%%%%%

\end{document}